\documentclass[%
 reprint,
nofootinbib,
 amsmath,amssymb,
 aps,
pra,
]{revtex4-2}

\usepackage{revtex}
\usepackage{comandi_tesi}
\usepackage{comandi_ndp}
\renewcommand{\disp}[1]{\hat{D}\left(#1\right)}

\begin{document}
\title{An agnostic-Dolinar receiver for coherent states classification}

\author{Fabio Zoratti}
\affiliation{Scuola Normale Superiore, I-56126 Pisa, Italy}
\author{Marco Fanizza}
\affiliation{NEST, Scuola Normale Superiore and Istituto Nanoscienze-CNR, I-56126 Pisa, Italy}
\author{Nicola Dalla Pozza}
\affiliation{Scuola Normale Superiore, I-56126 Pisa, Italy}
\author{Vittorio Giovannetti}
\affiliation{NEST, Scuola Normale Superiore and Istituto Nanoscienze-CNR, I-56126 Pisa, Italy}

\date{\today}

\begin{abstract}

  We consider the problem of discriminating quantum states, where the task is to distinguish two different quantum states with a complete classical knowledge about them, and the problem of classifying quantum states, where the task is to distinguish two classes of quantum states where no prior classical information is available but a finite number of physical copies of each classes are given. In the case the quantum states are represented by coherent states of light, we identify intermediate scenarios where partial prior information is available. We evaluate an analytical expression for the minimum error when the quantum states are opposite and a prior on the amplitudes is known. Such a threshold is attained by complex POVM that involve highly non-linear optical procedure. A suboptimal procedure that can be implemented with current technology is presented  that is based on a modification of the conventional Dolinar receiver. We study and compare the performance of the scheme under different  assumptions on the prior information available.

\end{abstract}

\maketitle

\section{Introduction}

Since its early studies, the discrimination of quantum states has been a central problem in quantum information theory due to the impossibility of perfectly distinguish non--orthogonal quantum states. Its implications reflects not only on quantum communication scenarios~\cite{Gisin2007, Gyongyosi2018, Manzalini2020}, but also in metrology~\cite{Giovannetti2004, Giovannetti2006, Giovannetti2011, Pirandola2019}, sensing~\cite{Degen2017,Gefen2019}, quantum key distribution and cryptography~\cite{Gisin2002, Lo2014,Banaszek1999, Pirandola2020, Cavaliere2020}.

In the typical \emph{discrimination} scenario~\cite{Chefles2000, Bergou2007, Bergou2010, Barnett2009}, two parties, the transmitter and the receiver, agree on a known shared communication protocol, which defines the (possibly finite) set of quantum states to transmit and discriminate in the best way possible~\cite{Helstrom1976, Holevo1972}. A different perspective on
the problem has been adopted in recent years, following the growing trend of machine learning studies~\cite{Wittek2014Classification, Gambs2008, Guta2010, Schuld2014, Sasaki2002, Hayashi2005, Hayashi2006,Schuld2014, Chen2017}. In the context of supervised learning, \emph{classification} problems aim at assigning a sample to one of the available classes, of which a description is not known, but multiple training samples are provided. Training samples could be copies of the quantum states to classify or other members of the family defining the classes. Classification problems are more general since the description provided by the communication protocol allows the generation of the training samples, therefore expressing a discrimination problem as a classification one. The latter is also more difficult since the classifier has to learn a description or a strategy for the discrimination from the (possibly noisy) training samples, in addition to performing the distinction.

Historically, the discrimination scenario has been investigated the most. Minimum error discrimination has been considered initially for two quantum states~\cite{Helstrom1976, Holevo1972}, where the optimal solution for the measurement operators assigning the estimate has been given in a closed-form. Optimality conditions for a bigger set of quantum states have been found~\cite{Yuen1975}, but the evaluation of the measurement operators and of the performance usually requires numerical procedures such as semi--definite programming~\cite{Eldar2003}. When the quantum states exhibit symmetries, such evaluation can be further simplified~\cite{Eldar2001, Eldar2004, DallaPozza2015}. Despite the advances in the field, in the case of the discrimination of optical states, physical realizations of the optimal receiver end are still an open problem, with the only exception of the Dolinar receiver for the discrimination of two coherent states~\cite{dolinar_originale,dolinar_rivisitato_3, dolinar_rivisitato_2,dolinar_rivisitato_1, dolinar_experimental, dolinar_experimental, Bartuskova2008}. Along with this scheme, other practical realizations of suboptimal receivers have been proposed for other sets of coherent states~\cite{Kato1999,DallaPozza2014,Muller2015, DaSilva2014,Namkung2018}.

Regarding the quantum classification problem, early works frame the same scenario under different names, such as quantum matching (see~\cite{Sasaki2002} and references within) or quantum state identification~\cite{Hayashi2005}, and programmable discrimination~\cite{Bergou2005}.
In the minimum error setting, solutions for the two--classes problem came first for pure states~\cite{Hayashi2005, Hayashi2006} and then for general qubit mixed states~\cite{Guta2010,Sentis2010,Fanizza2019}, in the asymptotic and limited--training--samples regimes. Following papers~\cite{Guta2010,Sentis2012,Sentis2013} have focused on the performance comparison between a joint (collective) measurement strategy involving both the training samples and the one to distinguish, versus an Estimate\&Discriminate strategy, where the training copies are used to estimate the classes of states, and the classical information extracted is used to setup the discrimination.
The latter strategy results to be suboptimal to the former. The unambiguous version of the classification problem considered in this paper has been addressed in \cite{Sedlak2007, Sedlak2009}, which provided a strategy based on interferometers and photodetectors, that has been demonstrated experimentally~\cite{Bartuskova2008}. Other results on programmable discriminators can be found in~\cite{Fiurasek2002, Fiurasek2004, Dusek2002, Bergou2006b, Zhang2006, He2007, Herzog2008, Akimoto2011, Colina2012, Zhou2014}.

In the field of quantum optics, discrimination and classification problems have been formulated for the reading of an optical memory. The advantage of using quantum states of light for the discrimination has been established in a series of papers \cite{Tan2008, Lloyd2008, Pirandola2011, Pirandola2011b, Lupo2013}, while
in~\cite{Sentis2015} the reading has been framed as a classification problem between the vacuum state and a coherent state with unknown parameter, later setup to be a Gaussian a priori distribution around a mean value. The asymptotic behaviour of the collective measurement strategy and the Estimate\&Discriminate one has been evaluated and compared, confirming that the former gives better performances. In the paper, it is conjectured and given some evidence that this holds also for \emph{non--Gaussian} Estimate\&Discriminate strategies.

Our work further investigates this comparison. We consider the simplest scenario concerning the classification of an unknown state belonging to one of two classes of coherent states that are assigned by
giving access to a certain number of training copies.
A practical scenario where this task could be relevant is an optical link through a stochastic channel where the attenuation is so unpredictable and random that over sufficiently long time intervals, the signal intensity can be assumed to be completely unknown, and the same holds for the added phase. In this context one may try to exploit the existence of stability
periods in the perturbations induced by the noise, to set communication protocols that consist in sending
 samples of the two types of training signals followed by the quantum state to classify.
To begin with we show that via simple linear optics, the problem we are facing can always be reduced to the special symmetric case
where the two classes of inputs differs only by the sign of the associated coherent amplitudes.
We hence evaluate the optimal bound for the probability of success of the classification task, under the assumption that the protocol to use is phase invariant. Such threshold can be explicitly calculated for any number of input copies of the training states, and for any given  prior distribution of the coherent amplitudes
that characterize them.  Unforturnately the POVM measurement that ensures the attainability of the
optimal bound relays on highly non-linear optical processes that are not feasible with current technology. In alternative we propose a modification of the conventional  Dolinar receiver~\cite{dolinar_originale} that we dub agnostic-Dolinar receiver, whose implementation is instead at reach with conventional quantum optical
procedures.  While being sub-optimal when
 employed with a finite number $n$ of training copies, the proposed scheme it is shown to
 saturate to the optimal bound in the asymptotic limit  $n\rightarrow \infty$. Most importantly, for all $n$,
  it yields a clear advantage when
 compared with respect to simple
 Estimate\&Discriminate strategies that involve estimations performed on a fraction of the training samples.

The paper is organized as follows. In \cref{sec:adaptive_discriminators} we introduce the problem and review the original Dolinar receiver.  \cref{sec:agnostic}  is the main section of the manuscript. Here we reduce the problem to a symmetric scenario, provide an optimal bound for the problem, and present our apparatus. We study its performance by comparing it with an Estimate\&Discriminate strategy based on a miscalibrated Dolinar scheme, and also compute the optimal error probability of the problem under different assumptions on the prior information available.

\section{Discrimination and Classification of coherent optical signals
}\label{sec:adaptive_discriminators}

This section is dedicated to set the problem, introduce the notation, and review some basic facts.

\subsection{Discrimination vs Classification}

The discrimination and classification of quantum states are  two distinct  primitives of
quantum information processing that find applications in a variety of different contexts.
Relaying on the error probability as cost function to evaluate their efficiencies~\cite{Helstrom1976}, we schematize these procedures in terms
of the following Minimum Error Discrimination (MED), and Minimum Error Classification (MEC)
problems:
\\

{\bf MED problem:--}  {\it Given a set of known quantum states $\{\hat{\rho}_k\}_{k=1}^K$ and probabilities $\{p_k\}_{k=1}^K, \sum_k p_k=1$ and an unknown quantum state $\hat{\rho} \in \{\hat{\rho}_k\}$ drawn from the set with probability $p \in \{p_k\}$, find the POVM measurement operators $\{\hat{\Pi}_k\}_{k=1}^K$ that allows to identify $\rho$ with minimum probability of error $P_e^{(\rm{MED})}$, or equivalently, with maximum probability of correct decision $P_c^{\rm{(MED)}}=1-P_e^{(\rm{MED})}$,}
	\begin{equation}
	P_c^{\rm{(MED)}} = \sum_{k=1}^K  p_k  \tr \left[\hat{\Pi}_k \hat{\rho}_k \right ].
	\label{def:P_cDiscrimination}
	\end{equation}

It is worth reminding that for the special case  with $K=2$ an explicit solution for the MED problem is provided by the Helstrom theorem according to which the maximum  value of $P_c^{\rm{(MED)}}$ is achieved
by a binary projective measurement associated with positive and negative part of the operator $p_1 \hat{\rho}_1 - p_2 \hat{\rho}_2$, leading to
the optimal expression
	\begin{equation}
	P_{c,\max}^{\rm{(MED)}} =\frac{1}{2} \left( 1 + \mbox{tr} | p_1 \hat{\rho}_1 - p_2 \hat{\rho}_2| \right) \;,
	\label{def:P_cDiscriminationexact}
	\end{equation}
	which we shall employ in the following as a benchmark for the efficiency of our schemes. \\

{\bf MEC problem:--}
{\it Given a training set of quantum states $\{\hat{\rho}_k\}_{k=1}^n$ and a set of labels $\{y(k)\  \vert\  y(k) \in [1, \ldots, K]\}_{k=1}^n$ that associate each sample to its class $y(k)$, and given an unknown testing set of quantum states $\{\hat{\rho}_r\}_{r=1}^s$ and labels $\{z(r)  \vert\  z(r) \in [1, \ldots, K]\}_{r=1}^s$, find the POVM measurement operators $\{\hat{\Pi}_z\}_{z=1}^K$ that identify their classes with minimum error probability $P_e^{\rm{(MEC)}}$, or equivalently, with maximum probability of correct decision $P_c^{\rm{(MEC)}}=1-P_e^{\rm{(MEC)}}$,}
	\begin{equation}
	P_c^{\rm{(MEC)}} = \frac{1}{s} \sum_{r=1}^s \tr \left[ \hat{\Pi}_{z(r)} \hat{\rho}_r \right ]\;.
	\label{def:P_cClassification}
	\end{equation}

Notice that at variance with the MED problem, for the MEC problem  a (complete classical) description of the quantum states of the training set $\{\hat{\rho}_k\}_{k=1}^n$ is not known, nor it is the a priori probability of each class which is often inferred from the relative amount of the labels. Notice also that, even though in the subsequent sections of our manuscript we shall not adopt such options,  in the general MEC setting i) the samples of the training set  are not necessarily organized into cluster of identical copies, and ii)  the testing samples $\{\hat{\rho}_r\}_{r=1}^s$ are not  included in the training set $\{\hat{\rho}_k\}_{k=1}^n$.

\subsection{Quantum optical setting} \label{sec:QOS}
In the rest of the paper we shall consider
the case where the system of interest is a single optimal mode  of the electro-magnetic field
described by the annihilation and creation operators $\hat{a}$, $\hat{a}^\dag$ fulfilling canonical commutation rules, $[\hat{a},\hat{a}^\dag]=1$~\cite{serafini_quantum_continuous_variables, milburn_qoptics}.
In this context we
focus on a MEC problem where the training set is formed by identical  copies of $K=2$
 coherent states
  \begin{eqnarray}
 |{\alpha_k}\rangle = \hat{D}(\alpha_k) | \O\rangle\;,  \qquad k \in \{ 1, 2\}\;,
 \end{eqnarray}
 whose complex amplitudes $\alpha_1$ and $\alpha_2$ are unknown (in this expression $| \O\rangle$  and $\hat{D}(\alpha) =
  \exp[ \alpha \hat{a}^\dag - \alpha^* \hat{a}]$ are respectively the vacuum state  and the displacement operator of the model).
  Specifically we shall work under the assumption of having $n$ copies of each training state, and  that
  the density matrix we need to classify is guaranteed to be a coherent state $\ket{\delta}$ that coincides either with $|\alpha_1\rangle$ or with $|\alpha_2\rangle$, with flat
  prior   probabilities  (i.e. $p_1=p_2=1/2$).
  The resulting  global input state we  operate can hence be expressed in the  following  multi-mode compact form
  \begin{equation}
    \label{INPUT1}
    \ket{\alpha_1^{\otimes n} ,\alpha_2^{\otimes n} ,\delta} = \ket{\alpha_1}^{\otimes n}\otimes \ket{\alpha_2}^{\otimes n}\otimes\ket{\delta}\;,
  \end{equation}
which in principle is characterized by 4 unknown real parameters  (the complex numbers $\alpha_1$
and $\alpha_2$), and by one quantum binary variable (the testing state $\ket{\delta} \in \{ | \alpha_1\rangle,| \alpha_2\rangle \}$).
In the MED version of the problem, i.e. when the complex quantities $\alpha_1,\alpha_2$ are assigned or equivalently when the number $n$ of the training copies of the MEC problem are infinitely many so that the values of
the amplitudes can be recovered through quantum process tomography, the optimal success probability
can be computed as in
\cref{def:P_cDiscriminationexact} leading to the value
	\begin{eqnarray}
	P_{c,\max}^{\rm{(MED)}}&=&\frac{1}{2} \left( 1 + \sqrt{1-4p_1p_2
	|\langle \alpha_1 |\alpha_2\rangle|^2} \right) \nonumber \\
	 &=&\frac{1}{2} \left( 1 + \sqrt{1-4p_1p_2  \; \ee^{-|\alpha_1-\alpha_2|^2}} \right) \;,
	\label{def:P_cDiscriminationexact1}
	\end{eqnarray}
	which can be attained via the Dolinar detection scheme which we review in the next section.
	The procedure we have in mind to solve the MEC problem for finite $n$ is a variation of the such scheme that
	relays on basic linear optical manipulations of the state $\ket{\alpha_1^{\otimes n} ,\alpha_2^{\otimes n} ,\delta}$ to compensate for the absence of classical information on the
values of the amplitudes $\alpha_1$ and $\alpha_2$. We call such procedure
agnostic-Dolinar receiver and we present it in \cref{sec:agnostic}.

\subsection{Dolinar receiver}\label{sec:dolinar_receiver}
As anticipated the Dolinar receiver  is an experimental technique that allows one to practically attain the
optimal threshold limit~(\ref{def:P_cDiscriminationexact1}) for a binary MED problem aimed to discriminate between two assigned coherent input states $|\alpha_1\rangle$ and $|\alpha_2\rangle$ which are produced with
prior probabilities $p_1$ and $p_2$.
It is worth noticing that in this special context, due to the fact that the values of the complex amplitudes $\alpha_1$ and $\alpha_2$ are known, one can always reduce the problem to case of a symmetric configuration in which $|\alpha_1\rangle$, $|\alpha_2\rangle$ are traded with  the couple  $|\pm \bar{\alpha}\rangle$ with $\bar{\alpha}=(\alpha_1-\alpha_2)/{2}$, or equivalently to a maximally
 energetically unbalanced setting where instead $|\alpha_1\rangle$ and  $|\alpha_2\rangle$ get replaced  by  $|2\bar{\alpha}\rangle$ and $|\O\rangle$, respectively. Such mappings in fact simply
 relay on acting on the input $|\delta\rangle$ via  optical displacements, i.e. transformations that can be physically implemented by
  mixing the signal with an intense coherent ancillary state through a  beam-splitter of high transmissivity. Specifically in the cases we are considering  this accounts in replacing  $|\delta\rangle$  with
  \begin{equation}\label{disp}
 \disp{-\beta} \ket{\delta} = e^{(\alpha \beta^* - \alpha^* \beta)/2}   \ket{\delta -\beta}\;,
 \end{equation}
 with $\beta= (\alpha_1+\alpha_2)/{2}$ for the symmetric configuration, and  $\beta= -\alpha_2$ for the maximally
 energetically unbalanced setting. In view of these facts  in the following paragraphs,  without loss of generality we shall  assume the symmetric setting posing $\alpha_1=-\alpha_2=\alpha$.
We also point out that in our presentation of the Dolinar scheme we shall rely to the continuous-time formulation of the problem
discussed in~\cite{dolinar_rivisitato_1} -- see however
Appendix~\ref{SEC:DOLINAREQ} for a description based on sequence of beam splitters and photon--detectors.
\\

\begin{figure}[t]
  \centering
  \includegraphics[width=0.43\textwidth]{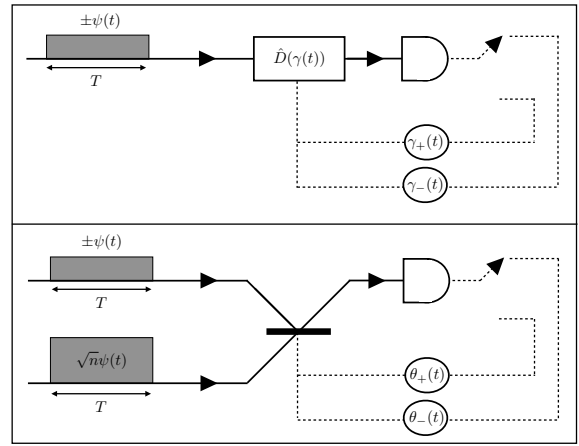}
  \caption{Time continuous description of the conventional Dolinar receiver (upper panel) and the agnostic-Dolinar receiver (lower panel): continuous black lines represent optical signals, dashed lines represent instead  classical control lines.
  In the upper panel, the rectangle with the $\disp{\gamma(t)}$ symbol represents a time-dependent displacement gate, which is followed by a photon-counter that switch between two classically controlled quantities represented in the circles. In the lower panel, the crossing with an extra horizontal rectangle is a beam splitter with time-dependent reflectivity $\theta_k(t) \in \{\theta_+(t), \theta_-(t)\}$, which are classically controlled and selected by the photon-counter.}\label{fig:dolinar_schema}
\end{figure}

The Dolinar receiver works by continuously applying a displacement $\hat{D}(\gamma_k(t))$ on the input state and performing a photon-counting on the displaced signal (see upper panel of \cref{fig:dolinar_schema}). The displacement is optimized such that the parity of the counting at the end of the signal gives the final estimate of the input coherent state.
The rationale of the scheme follows the same idea of another suboptimal scheme, the Kennedy receiver~\cite{kennedy_receiver}.
The key idea behind the Kennedy receiver is to rigidly shift the two states by $\alpha$ to obtain the pair $\{\ket{2\alpha}, \ket{0}\}$ (i.e. to map the symmetric setting into the maximally energetically unbalanced one), and then perform photon-counting. The vacuum always counts zero photons, and thus the Kennedy receiver uses $\hat{\Pi}_+ = \id - \ketbra{\O}{\O},\
\hat{\Pi}_-=\ketbra{\O}{\O}$ as the POVM for the discrimination. This apparatus does not reach the optimal Helstrom bound~(\ref{def:P_cDiscriminationexact1}), and this gap is filled exactly by the Dolinar receiver~\cite{dolinar_originale,dolinar_rivisitato_1,dolinar_rivisitato_2}, where the optimal shift is defined from an optimization procedure.
More precisely, consider the input field for our system $\psi_k(t),\ 0 < t < T$, corresponding to the coherent state $\ket{\pm \alpha}$, being represented by
\begin{align}
  \psi_k(t) = \pm \psi e^{i \omega_0 t},   \qquad (k=\pm)\;,
	\label{eq:input_field}
\end{align}
with $\omega_0$ the optical pulsation frequency and $T$ the pulse duration, that are linked through the normalization condition
\begin{align}
  \abs{\alpha}^2 = \dint_0^T \abs{\psi(t)}^2 \dd t = \abs{\psi_k}^2 T.
\end{align}
It can be shown that all the following results do not depend on our choice of $T$ and thus, for clarity reasons, we set $T= 1$ and $\psi = \alpha$. This input signal is displaced by a value $\gamma_k(t)$, which is classically controlled via feedback, and the resulting sum signal is monitored with a photon-counter. Every time our counter ``clicks'', the added signal is discontinuously changed from $\gamma_+(t)$ to $\gamma_-(t)$ and vice-versa, which are the displacement employed when the provisional estimation for the quantum state is $z(t)=+$ or $z(t)=-$ corresponding to the sign of $\ket{\pm \alpha}$. Equivalently, the parity of the total number of photons counted until the current time $t$ gives the estimation for the quantum state. The final discrimination result is declared at the end of the signal, $t = T = 1$.

We can use the following argument to find the optimal choice for $\gamma_k(t)$.
The provisional correct decision probability at time $t$ can be written as
\begin{align}
  P_c(t) = P[z(t) = + \vert k=+]p_+ + P[z(t) = - \vert k=-]p_-.
\end{align}
Let us now assume that $k=+$, that is, the actual coherent to discriminate is represented by the input field $\psi_+(t)$ of \cref{eq:input_field}. Then, the process $z(t)$ can be interpreted as a telegraph process driven by non-homogeneous Poisson processes~\cite{dolinar_rivisitato_1} with rates
\begin{align}
  \label{eq:lambda_mu_dolinar}
  \lambda(t) &= \abs{\psi_+(t) - \gamma_+(t)}^2, \nonumber \\
  \mu(t) &= \abs{\psi_+(t) - \gamma_-(t)}^2,
\end{align}
that allows the evaluation of the differential equations for the conditional probabilities of correct decision $q_+(t) = P[z(t)=+ \vert k=+]$ and $q_-(t) = P[z(t)=- \vert k=-]$ as
\begin{align}
  \frac{\dd q_+ (t)}{\dd t} &= \mu(t) - [\lambda(t) + \mu(t)] q_+(t), \nonumber \\
  \frac{\dd q_- (t)}{\dd t} &= \mu(t) - [\lambda(t) + \mu(t)] q_-(t).
	\label{eq:differential_equations}
\end{align}
Hence, the differential equation for the correct detection  probability results
\begin{align}
  \label{eq:differenziale_P_c}
  \frac{\dd P_c(t)}{\dd t} = \frac{q_+' + q_-'}{2} = \mu(t) - [\lambda(t) + \mu(t)] P_c(t)\;.
\end{align}
We can now extremize with respect to $\gamma_+(t) = -\gamma_-(t)$ at each fixed time to find the optimal displacement, which leads to the differential equation
\begin{align}
\frac{\dd P_c(t)}{\dd t} & = -4|\alpha|^2 \frac{P_c(t)(1-P_c(t))}{1-2P_c(t)} \notag \\
& = |\alpha|^2 \left( 1-2P_c(t) - \frac{1}{1-2P_c(t)}  \right )\label{eq:optimal_differenziale_P_c}
\end{align}
with solution
\begin{equation}
\label{eq:Dolinar_P_c}
P_{c,\text{Dol}}^{\rm{(MED)}}(t) = \frac{1}{2} \left( 1 + \sqrt{1-4p_+p_- \ee^{-4|\alpha|^2 t}} \right)\;,
\end{equation}
which for $t=1$ reaches the maximum value~(\ref{def:P_cDiscriminationexact1}) dictated by the
Helstrom bound~\cite{Helstrom1976}.
 An experimental realization of this apparatus has been realized in~\cite{dolinar_experimental}.

\section{Building an agnostic-Dolinar Receiver} \label{sec:agnostic}

In this section we present a scheme that allows one to solve the MEC problem associated with a binary
set of coherent inputs introduced in \cref{sec:QOS}. As a preliminary step, following the discussion at beginning
of \cref{sec:dolinar_receiver},  in \cref{sec:simp1}  we show that via some trivial physical manipulations of the input data we
 can always restrict the analysis to the special case of a binary MEC problem where $\alpha_2=-\alpha_1$, hence reducing  from 4 to 2 the
 number of unknown real parameters associated with the input state \cref{INPUT1}.
 In \cref{sec:conti_analitici_fanizza} we find the equivalent of the Helstrom bound in \cref{def:P_cDiscriminationexact1} for the MEC scenario:  as we shall see the attainability of such optimal threshold relay on
 the possibility of implementing highly non-linear optical processes which represent an impressive challenge for current technology.
In \cref{sec:dolinar_agnostico} we hence focus on a more realistic procedure based on
 an adaptive scheme where
 the displacements $\hat{D}(\gamma_k(t))$ of  the original Dolinar receiver are replaced by
 partial coherent mixing  with a fraction of the copies of the training set.
 In this context we show that if the value  $|\alpha_1-\alpha_2|$ is known (an assumption which would be trivially granted in the MED version of the problem but not in the MEC scenario where it allows to reduce the number of the unknown real parameters needed from 2 to 1), the new setting  attains a probability of success that already for medium size  values of $n$, approaches the one of the non-linear optimal bound of
 \cref{sec:conti_analitici_fanizza}.
In Sec.~\ref{sec:dolinar_agnostico_performance} we fix the issue associated with the lack of knowledge of the parameter  $|\alpha_1-\alpha_2|$ by exploiting  part of the copies of the training set to obtain a preliminary estimation of such term:
the performance of the resulting scheme is hence studied and compared with those one would obtain by using a
mis-calibrated Dolinar scheme.

\subsection{Mapping the MEC problem into a symmetric scenario} \label{sec:simp1}

Aim of this section is to show that when studying the MEC problem introduced in Sec.~\ref{sec:QOS}, we can safely assume the amplitudes of the unknown coherent states of the training set, to  have opposite phases and equal
absolute values (i.e.
$\alpha_2=-\alpha_1$). This simplification is analogous to the reduction of the general  MED problem to a symmetric configuration: in the present case however this formal passage
 is slightly more subtle due
to the fact we do not have prior classical knowledge of  $\alpha_1$ and $\alpha_2$.

 A key ingredient  of the  analysis is represented by what we may call a $m$-modes {\it concentrator gate} $\hat{U}_{C}^{(m)}$~\cite{Sentis2015,Sedlak2009,guha_2011_optical_receivers,sedlak2008unambiguous}, i.e. a  $m$-modes unitary transformation implementable via array of properly concatenated beam-splitters, that
 acting on a collection of
 $m$ copies of a generic (possibly unknown) coherent state $|\alpha\rangle$ manages to move all their photons in a single output mode via the mapping
 \begin{equation} \label{concGATE}
 |\alpha\rangle^{\otimes m} \longmapsto \hat{U}_{C}^{(m)} |\alpha\rangle^{\otimes m} =  |\sqrt{m} \alpha\rangle \otimes |\O\rangle^{\otimes m-1} \;.
 \end{equation}
Applying this to
the $(2n+1)$-modes input state~(\ref{INPUT1}) that formally defines
the MEC problem we are facing, we can map it into an
 equivalent form where all photons are concentrated into the following  3-modes  coherent state
 \begin{equation}
   \ket{\sqrt{n}\alpha_1,\sqrt{n} \alpha_2 ,\delta}=  \ket{\sqrt{n} \alpha_1}  \otimes \ket{\sqrt{n}\alpha_2}\otimes \ket{\delta}\;, \label{NEWINPUT1}
 \end{equation}
 (the net operation involves
 a collection of extra $2(n-1)$ irrelevant vacuum  states $\ket{\O}$).
Notice also that  with a 3--port beam splitter~\cite{serafini_quantum_continuous_variables} defined by the $3\times 3$ scattering matrix
  \begin{align}
    \label{eq:3beam_splitter_dolinar_generale}
    S_n =
  \begin{bmatrix}
     \frac{1}{\sqrt{2}}  &  - \frac{1}{\sqrt{2}} & 0\\
    \frac{1}{\sqrt{4n + 2}} & \frac{1}{\sqrt{4n + 2}}& -\sqrt{\frac{2n}{2n + 1}} \\
    \sqrt{\frac{n}{2n + 1}}  &  \sqrt{\frac{n}{2n + 1}} & \sqrt{\frac{1}{2n + 1}}
  \end{bmatrix}\;,
  \end{align}
we can then unitarily transform (\ref{NEWINPUT1}) (and hence (\ref{INPUT1})) into the further equivalent  form
 \begin{equation} \ket{\sqrt{2n+1}\alpha'} \otimes \ket{\delta'} \otimes\ket{\tfrac{\delta + n(\alpha_1 + \alpha_2)}{\sqrt{2n + 1}}} \;,  \label{NEWNEWINPUT1}
 \end{equation}
 where now
 \begin{equation} \alpha'  = \sqrt{\tfrac{2n}{2n + 1}} \tfrac{\alpha_1 - \alpha_2}{2}\;,\end{equation}   and where
 \begin{equation} \delta' = \sqrt{\tfrac{2n}{2n + 1}}\left(\tfrac{\alpha_1 + \alpha_2}{2}-\delta\right)\;, \end{equation}
is a variable that for $\delta = \alpha_1,\alpha_2$
assumes  the values $\pm \alpha'$.
Therefore, since the coherent state  $\ket{\sqrt{2n+1}\alpha'}$ can be mapped into $ \ket{\alpha'}^{\otimes (2n +1)}$
via the action of  the inverse of a concentrator gate $\hat{U}_{C}^{(2n+1)}$
our original   MEC problem associated with the input (\ref{INPUT1}), can be casted into the new MEC
problem where starting from a collection of $2n+1$ copies of the coherent state $|\alpha'\rangle$ we
are asked to decide wether the state $|\delta'\rangle$ is either equal to $|\alpha'\rangle$ or to $|- \alpha'\rangle$. Notice that in doing so we are implicitly  neglecting  the last coherent state component  of \cref{NEWNEWINPUT1}: this however does not represent a huge loss of information since the residual  dependence that such term  bares upon  $\delta$ vanishes  in the asymptotic limit of $n\rightarrow \infty$, and in any case the analysis shows that a scheme that is capable to efficiently solve a MEC  in the symmetric scenario can also be applied to the
generic one. Notice further that via the action of a phase shifter gate aimed to flip the sign of the
amplitude of an incoming input state, we can also convert the $2n+1$ copies of $|\alpha'\rangle$
into a state of the form $|\alpha'\rangle^{\otimes m}\otimes |\alpha'\rangle^{\otimes 2n+1-m}$ with $0\leq m\leq 2n+1$.
Relying on these observations in  the remainder of the paper
we will  thus focus on the symmetric version of our MEC problem where  starting from the beginning it is  assumed $\alpha_1=-\alpha_2=\alpha$, hence replacing the input state~(\ref{INPUT1}) with the vector
 \begin{equation}
\label{INPUT1sym}  \ket{\alpha^{\otimes n},\delta} = \ket{\alpha}^{\otimes  n}\otimes\ket{\delta}\;,\end{equation}
characterized by 2 unknown real parameters  (the phase and the absolute values of the complex number $\alpha$), and by the quantum binary variable  $|\delta\rangle\in\{ |\pm\alpha\rangle \}$
[NB. formally speaking in the above expression the total number of copies of $|\alpha\rangle$ we
can extract from (\ref{INPUT1}) requiring  $\alpha_1=-\alpha_2$ would be $2n$: hereafter however we shall reparametrize this with $n$ just to allow for the possibility of having  an odd number of input copies].

\subsection{Optimal bound for the problem}\label{sec:conti_analitici_fanizza}

As already anticipated
in the asymptotic limit
 $n \gg 1$, the optimal upper  bound for the success probability of a generic apparatus
 aimed to solve the MEC problem associated with the input~(\ref{INPUT1sym}) reduces to the Helstrom limit~(\ref{def:P_cDiscriminationexact1})
 attainable via the Dolinar scheme, i.e. the quantity
 \begin{eqnarray}
	P_{c,\max}^{\rm{(MED)}}&=&\frac{1}{2}  \left( 1 + \sqrt{1- \ee^{-4|\alpha|^2}} \right) \;.
	\label{def:P_cDiscriminationexact1alpha}
	\end{eqnarray}
Estimating the  optimal MED performance in the finite-copy case
can be useful to compare our results with a fundamental bound. In this section we will find this bound
under the assumption that the protocol we use is phase invariant, i.e. insensitive to the phase value of
the amplitude $\alpha$ that enters in Eq.~(\ref{INPUT1sym}), a constraint which is reasonable to impose
in the MED scenario where no prior info on $\alpha$ is granted.
For this purpose, first of all  we invoke once more the action of a concentrator gate~(\ref{concGATE}),
to replace  $\ket{\alpha^{\otimes n},\delta}$ with a two mode input state
$\ket{\sqrt{n} \alpha,\delta}$.
Then we focus on two-element POVM $\{\hat{E}_+,\hat{E}_-=\hat{I}
 -\hat{E}_+\}$ which, acting globally on the two modes of the model, aims to discriminate  the density matrix $\hat{\rho}_+=\ket{\sqrt{n} \alpha,\alpha}\bra{\sqrt{n} \alpha,\alpha}$ from $\hat{\rho}_-=\ket{\sqrt{n} \alpha,-\alpha}\bra{\sqrt{n} \alpha,-\alpha}$ under phase invariant assumptions.
 It is worth stressing that a similar calculation was performed in Ref.~\cite{Sentis2015} for a slightly different
 setting where the two states under scrutiny were  $\ket{\sqrt{n} \alpha,\alpha}$ and $\ket{\sqrt{n} \alpha,0}$ and where the analysis was confined in large $n$ limit:
 as we shall see in the following, at variance with those results, due to the symmetric structure of the inputs
 we employ,
 our analysis allows us to present closed analytical expressions also for the finite $n$ limit.
Specifically we associate  $\hat{E}_+$ to $\hat{\rho}_+$ and $\hat{E}_-$ to $\hat{\rho}_-$,
and we enforce the phase invariant constraint by requiring them to commute with the global phase operator
$e^{i \phi (\hat{n}_1+\hat{n}_2)}$ with $\hat{n}_1=\hat{a}^\dag_1 \hat{a}_1$ and $\hat{n}_2=\hat{a}^\dag_2 \hat{a}_2$ being the number operators of the two modes of the model.
By Schur's lemma~\cite{hall2003lie} it then follows that the POVM elements must satisfy the identities $\hat{E}_\pm=\sum_{m=0}^{\infty}\hat{\Pi}_m\hat{E}_\pm \hat{\Pi}_m=\sum_{m=0}^{\infty}\hat{E}_{+,m}$, where $\hat{\Pi}_{m}$ is the projector on the subspace total photon number $n_1+n_2=m$ and $\{\hat{E}_{+,m},\hat{E}_{-,m}=\hat{I}_m-\hat{E}_{+,m}\}$ is a binary POVM in this subspace.
For any two states $\rrho_+$ and $\rrho_-$, POVMs $\{\hat{E}_{\pm}\}$ with this property, the probability of error satisfies
\begin{align}
P_e &= \frac{1}{2}\left(1-\sum_{m=0}^{\infty} \frac{\tr[\hat{E}_-\hat{\Pi}_m \rrho_+ \hat{\Pi}_m+ \hat{E}_+\hat{\Pi}_m \rrho_-\hat{\Pi}_m]}{2}\right) \nonumber
\\
\label{MINIMO} &\geq \frac{1}{2}\left(1-\sum_{m=0}^{\infty} \frac{\tracenorm{\hat{\Pi}_m(\rrho_+-\rrho_-)\hat{\Pi}_m}}{2}\right)\;.
\end{align}
where the inequality comes from the Helstrom bound.
Remembering that in our case $\rrho_\pm$ are the coherent states $\ketbra{\sqrt{n}\alpha}{\sqrt{n}\alpha} \otimes \ketbra{\pm\alpha}{\pm\alpha}$
 it follows that the optimal choice for $\hat{E}_{\pm}$ is given by
\begin{eqnarray}  \label{OPTMIN}
\hat{E}_{\pm} := \sum_{m=0}^{\infty} | m; \pm \rangle\langle m; \pm| \;,
\end{eqnarray}
with
\begin{equation} \label{defMp}
\ket{m,\pm}:=\sum_{n_1+n_2=m}\sqrt{\binom{m}{n_1}}\frac{\sqrt{n}^{n_1}(\pm 1)^{n_2}}{\sqrt{n+1}^{m}} \ket{n_1,n_2}\;.
\end{equation}
Accordingly as shown in \cref{sec:conti_fanizza_analitici_appendice}, the minimum error probability (\cref{MINIMO}) reduces to
\begin{align}\label{NUOVAnew}
P_{e,\min}^{({\rm MEC},n)} & = \frac{1}{2}\left(1-\frac{1}{2}\sum_{m=0}^{\infty} \mathfrak{p}(m; \mu)\sqrt{1-\left(\tfrac{N-1}{N+1}\right)^{2m}}\right) \;, \end{align}
where  $\mu= \sqrt{n+1}\; \alpha$ and
$\mathfrak{p}(m; \mu)$ is the probability of drawing $m$ from a poissonian of mean
$|\mu|^2$, namely
\begin{align} \label{eq:optimal_error_phase_invariant}
  \mathfrak{p}(m; \mu) = \frac{\abs{\mu}^{2m}}{m!} \exp[-\abs{\mu}^2].
\end{align}
Notice in particular that, for fixed $|\alpha|^2$, the above expression admits the following asymptotic expansion
at large~$n$
\begin{align}
P_{e,\min}^{({\rm MEC},n)}  \simeq
&\frac{1}{2}\left[1-\frac{1}{2}\left(\sqrt{1-e^{-4|\alpha|^2}}-\frac{1}{n} \tfrac{2|\alpha|^2e^{-4|\alpha|^2}}{\left(1-e^{-4|\alpha|^2}\right)^{3/2}}\right)\right]\nonumber\\&+O\left(\tfrac{1}{n^2}\right),
\end{align}
with the leading term corresponding to value dictated by the Helstrom limit~(\ref{def:P_cDiscriminationexact1alpha}).
The same analysis can be repeated for input states of our problem with a given prior $p(\alpha)$ by simply
replacing $\rrho_\pm$ with
\begin{equation}
  \label{eq:mixed_states_phase_invariant}
\rrho\aap{(ave)}_{\pm}=\int_{\mathbb C^2} \dd\alpha \;  p(\alpha) \ketbra{\sqrt{n}\alpha}{\sqrt{n}\alpha} \otimes \ketbra{\pm\alpha}{\pm\alpha}\;,
\end{equation}
obtaining in this case the following optimal minimum error probability
\begin{align}
\bar{P}_{e,\min}^{({\rm MEC},n)} = \frac{1}{2}\left(1-\frac{1}{2}\sum_{m = 0}^\infty \bar{\mathfrak p}(m)\sqrt{1-\left(\frac{n-1}{n+1}\right)^{2m}}\right)\;,
        \label{eq:error_phase_invariant}
\end{align}
where now
\begin{eqnarray}
\bar{\mathfrak p}(m)=  \int_{\mathbb C^2} \dd \alpha  \; p(\alpha)\; \mathfrak{p}(m;  \sqrt{n+1} \alpha)\;,
\end{eqnarray}
is the average photon number distribution.

\subsection{Agnostic-Dolinar receiver with prior information on the  input mean photon number}\label{sec:dolinar_agnostico}

The implementation of the optimal covariant measure~(\ref{OPTMIN}) is highly non trivial as it requires
to discriminate between  non-orthogonal states $\ket{m,\pm}$ that involve complex superposition of two-mode
Fock states (see Eq.~(\ref{defMp})). To compensate for this
here we introduce a preliminary version of the agnostic-Dolinar scheme that
 assumes that only the phase of the complex parameter $\alpha$ of Eq.~(\ref{INPUT1sym}) to be unknown, but grants full knowledge about the mean photon number $|\alpha|^2$ of the inputs. In other words  we interpolate between the symmetric version of the MEC problem
 defined  at the end of the previous section (2 unknown real terms and one quantum binary variable), and the corresponding MED problem (zero unknown
 real terms and one quantum binary variable).
As schematically shown in the lower panel of  \cref{fig:dolinar_schema}, the basic idea is to replace
the displacement operations of the original Dolinar configuration whose values
$\gamma_k$  assume full knowledge of $\alpha$,
	 with coherent mixing of the testing input with
	the single-mode concentrated version $\ket{\sqrt{n}\alpha}$ of the training copies, via a beam-splitter operation
characterized by a time--dependent reflectivity  $\theta_k(t)$, where $k$ can be either $+$ or $-$. Because of the symmetry of the problem, we will call $\theta(t) = \theta_+(t) = \theta_-(t) + \pi$.
Note that in the alternative formulation of the Dolinar receiver~\cite{dolinar_rivisitato_2} discussed in Appendix~\ref{SEC:DOLINAREQ}  where the input coherent state is sliced by the sequence of beam splitter and on each slice a displacement and photon--counting process is applied, the scheme of the current section substitute the displacement operations with additional beam splitters, which mix slices of the $n$ training copies and the sliced input field in an optimized way to give output rates~\cref{eq:lambda_mu_beam_splitter}.
It is finally worth stressing that,
since we not relay on reference signals,
 by construction, the proposed detection  strategy is explicitly phase insensitive: accordingly the optimal bound~(\ref{NUOVAnew})  constitute
 a proper reference for testing the efficiency of the scheme.

 In order to evaluate the optimal $\theta(t)$ and maximize the correct decision probability we follow the same procedure of~\cref{sec:dolinar_receiver}, that we will now discuss with a more detail. First of all, following the same procedure, we define our new $\lambda(t), \mu(t)$ as

\begin{align}
\label{eq:lambda_mu_beam_splitter}
  \lambda(t) &=|\alpha|^2 \abs{\cos \theta(t) - \sqrt{n}\sin \theta(t) }^2, \nonumber \\
  \mu(t) &= |\alpha|^2  \abs{ \cos \theta(t) + \sqrt{n} \sin \theta(t)}^2.
\end{align}

\begin{figure}[t]
  \centering
    \begin{tikzpicture}
      \begin{axis}[
        xlabel={$n$},
        y label style={rotate=-90},
        ylabel={$Pe^*$},
        xmode=log,
        ymode=log,
        line width=.9pt,
        transpose legend,
        legend columns=2,
        legend style={at={(0.5,-0.15)},anchor=north},
        cycle multi list={%
          color list\nextlist
          [2 of]mark list
        }]
        ]
        \addplot[color=black, dashed]
        table [x=n, y=0.25, col sep=comma] {plot_differenziale_dolinar.txt};
        \addlegendentry{$\abs{\alpha} = 1/4$}

        \addplot[color=black, style=dotted]
        table [x=n, y=opt0.25, col sep=comma] {plot_differenziale_dolinar.txt};
        \addlegendentry{opt $\abs{\alpha} = 1/4$}

        \addplot[color=red, style=solid, forget plot]
        table [x=n, y=hel0.25, col sep=comma] {plot_differenziale_dolinar.txt};

        \addplot[color=black, dash pattern=on 1pt off 3pt on 3pt off 3pt]
        table [x=n, y=0.625, col sep=comma] {plot_differenziale_dolinar.txt};
        \addlegendentry{$\abs{\alpha} = 5/8$}

        \addplot[color=black, style=dotted]
        table [x=n, y=opt0.625, col sep=comma] {plot_differenziale_dolinar.txt};
        \addlegendentry{opt $\abs{\alpha} = 5/8$}

        \addplot[color=red, style=solid]
        table [x=n, y=hel0.625, col sep=comma, forget plot] {plot_differenziale_dolinar.txt};
      \end{axis}
    \end{tikzpicture}
  \caption{Error probability $P_e^*= 1-P^*_c(t=1)$ of an agnostic-Dolinar receiver (dashed and dash--dotted lines) as a function of the number $n$ of training copies, for different (known) values of the mean photon number  $|\alpha|^2$. The horizontal solid red lines correspond to the Helstrom
 values~(\ref{def:P_cDiscriminationexact1alpha})
 attained by the conventional Dolinar scheme
  for the associated MED problem. The dotted lines correspond to the optimal error probability obtainable  using the optimal bound for phase invariant scheme of~\cref{NUOVAnew}.}\label{fig:andamento_n_decision_prob}
\end{figure}
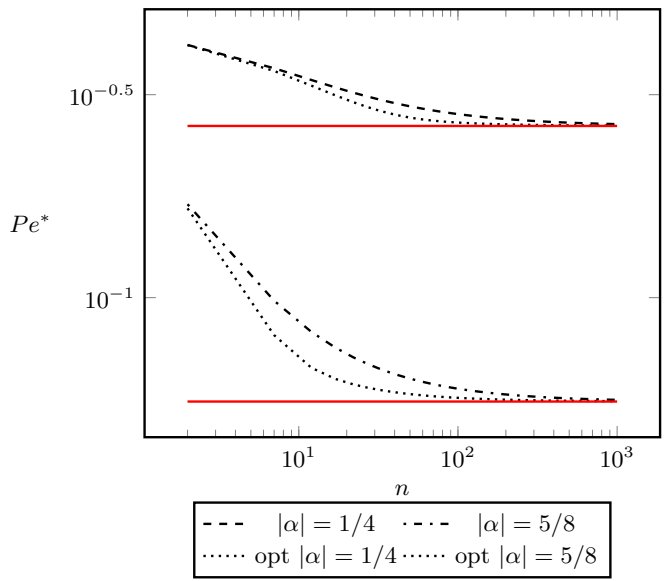

Let us now assume that $k=+$. The number of photons counted in an interval $(t,t+\Delta t]$ is a Poisson variable $N(t,t+\Delta t)$ with parameter $\lambda(t) \Delta t$ or $\mu(t) \Delta t$ depending whether the provisional hypothesis is respectively $z(t)=+$ or $z(t)=-$. These rates allows us to evaluate the conditional probabilities of correct decision $q_+(t) = P[z(t)=+ \vert k=+]$ and $q_-(t) = P[z(t)=- \vert k=-]$ following the Dolinar receiver strategy, which changes the provisional hypothesis when a photon is detected~\cite{dolinar_rivisitato_1}. From the difference equation
\begin{align}
& q_+(t+\Delta t) \nonumber \\
& \enspace = P[N(t, t+\Delta t)=0, z(t)=+ \vert k=+] + \nonumber \\
& \quad P[N(t, t+\Delta t)=1, z(t)=- \vert k=+] + o(\Delta t) \nonumber \\
& \enspace = P[N(t, t+\Delta t)=0 \vert z(t)=+, k=+] q_+(t) + \nonumber \\
& \quad P[N(t, t+\Delta t)=1 \vert z(t)=-, k=+] (1-q_+(t)) + o(\Delta t) \nonumber \\
& \enspace = (1-\lambda(t)\Delta t) q_+(t) + \mu(t) \Delta t (1-q_+(t)) + o(\Delta t),
\end{align}
follows the differential equation
$$
\frac{\dd q_+ (t)}{\dd t} = \mu(t) - [\lambda(t) + \mu(t)] q_+(t).
$$
In a similar fashion, from the expression for $q_-(t+\Delta t)$ and employing symmetric arguments for the displacements, we get the other differential equation of \cref{eq:differential_equations}.
 With this same procedure, it is obvious that the equation for $P_c(t)$ maintains the same form of \cref{eq:differenziale_P_c}, i.e.
\begin{align}
  \label{eq:differenziale_P_c1}
  \frac{\dd P_c(t)}{\dd t}= \mu(t) - [\lambda(t) + \mu(t)] P_c(t)\;,
\end{align}
where now however the terms $\mu(t)$ and $\lambda(t)$ are defined in \cref{eq:lambda_mu_beam_splitter} instead of \cref{eq:lambda_mu_dolinar}
We can now extremize \cref{eq:differenziale_P_c1}  to obtain an equation for the optimal control function  $\theta^*(t)=\theta\ped{opt}^{\abs{{\alpha}}^2, (n)}(t)$ given $\abs{\alpha}^2$ and $n$: this yields the solution
\begin{align}
  \label{eq:optimality_condition_agnostic}
  \tan(2\theta^*(t))  &= \frac{\sqrt{n}}{n - 1} \frac{1}{P^*_c(t) - \frac{1}{2}},
\end{align}
where $P^*_c(t)=P_{c,\rm{opt}}^{\abs{{\alpha}}^2, (n)}(t)$
is the  the associated optimal probability of success
that can be obtained by
solving a differential equation which is more concisely expressed with the change of variable $\xi(t) = P^*_c(t)- 1/2$, i.e.
\begin{align}
  \label{eq:differenziale_P_c_semplificata}
  	&\frac{\dd \xi(t)}{\dd t} = \abs{\alpha}^2\left(-\xi(t) (n + 1) + \sqrt{(n - 1)^2 \xi^2(t) + n}\right) \\
		& = \abs{\alpha}^2 \left[-2\xi(t) -(n-1)\left( \xi(t) - \sqrt{\frac{n^2}{(n-1)^2}+\xi(t)^2} \right) \right], \notag
\end{align}
(notice the explicit dependence upon $\abs{\alpha}^2$) which for $n~\to~\infty$ converges to \cref{eq:optimal_differenziale_P_c}.
With the separation of variables we can formally integrate (\ref{eq:differenziale_P_c_semplificata}) obtaining
\begin{align}
  \label{eq:analytic_solution_inverse}
  \frac{1}{2} \abs{\alpha}^2 t &= -\frac{n - 1}{4n} \tanh ^{-1}\left(\tfrac{(n-1) \xi(t)}{\sqrt{(n-1)^2 \xi^2(t)+n}}\right)  \nonumber \\
  + \frac{n+1}{8 n} & \Big[ \tanh ^{-1}\left(\tfrac{2\xi(t)(n + 1)\sqrt{(n - 1)^2 \xi^2(t)+ n}}{2(n^2 + 2)\xi^2(t) + n}\right)  \nonumber \\
  &- \log \left(1-4 \xi^2(t)\right)\Big]\;.
\end{align}
This expression cannot be inverted explicitly but
it can be evaluated numerically. It turns out that the resulting
$P^*_c(t)$ does not coincide with the  optimal bound of~\cref{NUOVAnew}. Still it remains close to such function being increasing in $n$ and asymptotically reaching
  the  performance of the Dolinar scheme given in Eq.~(\ref{def:P_cDiscriminationexact1alpha}).
 This is explicitly shown in~\cref{fig:andamento_n_decision_prob}, where we plot the resulting associated
probability of error
$P_e^* = 1-P^*_c(t=1)$ as a function of the training set $n$ for a known value of $\abs{\alpha}^2$:
notice that the asymptotic  limit~(\ref{def:P_cDiscriminationexact1alpha}) is reached quickly for $n \approx 20$ even in the full quantum limit ($\abs{\alpha} < 1$).

\subsection{Agnostic-Dolinar receiver with no prior information on the  input mean photon number}\label{sec:dolinar_agnostico_performance}

The scheme of the previous Section requires the exact knowledge of  $\abs{\alpha}^2$, which is not granted in the original MEC problem. Here, we compensate for such lack of information by  splitting the $n$ training copies of $|\alpha\rangle$ into two sets, one of size $m$ used to obtain an estimate of the value of $\abs{\alpha}^2$, and the other set of size $n - m$ copies to realize the apparatus described in the former section.
Obviously, \emph{a priori} there is no optimal choice for the size of the two sets, as there is a trade--off between a good parameter estimate and the performance of the apparatus. Studying the optimal way to split our sets will be the aim of this section.
To estimate the classical value of $\abs{\alpha}^2$ we examine two strategies, photon counting and heterodyne detection.
In the former case, the outcomes of the measurement are discrete values $k \in \mathbb{N}$ associated with the count of photons of the state $|\alpha\rangle$ which get distributed according to the poissonian probability $\mathfrak{p}(k; \alpha)$ defined as in  Eq.~(\ref{eq:optimal_error_phase_invariant}).
When applied to a coherent state $\ket{\sqrt{m}\alpha}$, we can obtain an estimate $\frac{k}{m}$ for $\abs{\alpha}^2$. Note that due to the discrete nature of the outcomes, such estimate comes in discrete steps.
Heterodyne detection is obtained by mixing the coherent state $|\alpha\rangle$ with a strong local oscillator with higher optical frequency~\cite{serafini_quantum_continuous_variables}, and then measuring both quadratures. The measurement outcomes are continuous and can be represented with a complex value $\beta \in \mathbb{C}$ obtained with probability
\begin{equation}
P[\beta; \alpha] = \frac{\ee^{-\abs{\beta-\alpha}^2}}{\pi}\;.
\end{equation}
The estimate for $\abs{\alpha}^2$ can be obtained from the absolute value $\abs{\beta}^2$ of the complex outcome, which is obtained with probability
\begin{align} \nonumber
P[\abs{\beta}^2; \alpha] & = \int_0^{2\pi} \frac{\ee^{-(\abs{\alpha}^2 + \abs{\beta}^2 -2 \abs{\alpha \beta} \cos \phi})}{\pi} \abs{\beta} \dd \phi \\
& = 2\abs{\beta} \ee^{-\abs{\alpha}^2 - \abs{\beta}^2} I_0(2\abs{\alpha \beta})\;,
\end{align}
with $I_0(\cdot)$ the modified Bessel function of the first kind.

To obtain the real performance of the apparatus, we take the expectation value over this probability distribution of the performance of the apparatus using the \emph{wrong} estimate for  $|\alpha|^2$. Namely, we use \cref{eq:optimality_condition_agnostic} and \cref{eq:analytic_solution_inverse} with the amplitude $\abs{\tilde{\alpha}}^2$ estimated from $\ket{\sqrt{m}\alpha}$ to obtain $\theta\ped{opt}^{\abs{\tilde{\alpha}}^2, (n-m)}(t)$. Averaging the performance for all the estimates gives the probability of correct decision as a function of $\alpha,\ n,\ m$.

We compare the performance with the Helstrom bound~(\ref{def:P_cDiscriminationexact1}), but also with an  Estimate\&Discriminate procedure which assumes of using all the $n$ copies to obtain a classical estimate of $\alpha$ ($m=n$) and then apply the original Dolinar receiver. The performance of this straightforward method (which we dub miscalibrated Dolinar)  is studied in the next section, \cref{sec:dolinar_gaussiano}, while the performances of the agnostic Dolinar are studied in \cref{sec:dolinar_copie_finite_stima} and \cref{sec:prior}.
\\
\subsubsection{Estimate\&Discriminate scheme based on Miscalibrated Dolinar receiver}\label{sec:dolinar_gaussiano}

In the original setting of the Dolinar receiver, the value of $\alpha$ that uniquely determines the coherent state $\ket{\alpha}$ was known with arbitrary precision.
In the Estimate\&Discriminate approach we analyze here, the idea is to use all the $n$ copies of $|\alpha\rangle$ of the state (\ref{INPUT1sym}) to get
an estimate $\beta$ of $\alpha$  and then use this to build up
the Dolinar procedure.

\begin{figure}[t]
  \centering
  \includegraphics[width=0.49\textwidth]{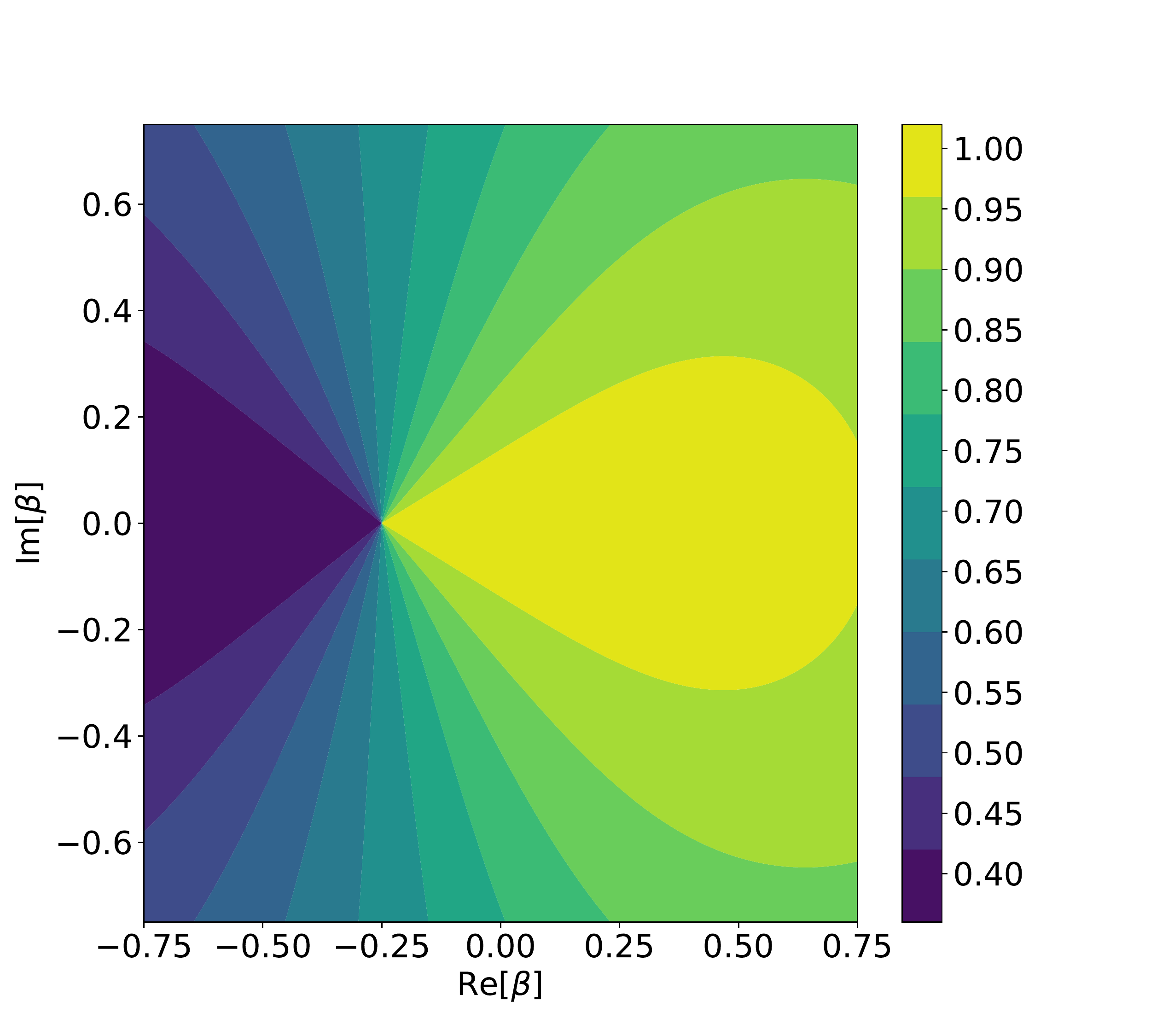}
    \caption{Density plot of the ratio between the probability of correct decision of a miscalibrated Dolinar receiver as in \cref{eq:P_c_Miscalibrated_Dolinar_receiver} and the optimal threshold $P_{c,\max}^{\rm{(MED)}}$ of \cref{def:P_cDiscriminationexact1alpha}
     as a function of the complex estimate $\beta$ for $\alpha=0.25$.}\label{fig:P_c_Dolinar_Scalibrato}
\end{figure}

\begin{figure}[t]
  \centering
  \begin{tikzpicture}
    \begin{axis}[
      xlabel={$n$},
      y label style={rotate=-90},
      ylabel={$Pe$},
      ymode=log,
      line width=.9pt,
      transpose legend,
      legend columns=2,
      legend style={at={(0.5,-0.15)},anchor=north},
      cycle multi list={%
        color list\nextlist
        [2 of]mark list
      }]

      grid style=dashed
      ]
      \addplot[color=black, solid]
      table [x=n, y=0.25, col sep=comma] {plot_helstrom_heterodyne.txt};
      \addlegendentry{$\alpha = 1/4$}

      \addplot[color=red, style=solid, forget plot]
      table [x=n, y=hel0.25, col sep=comma] {plot_helstrom_heterodyne.txt};

      \addplot[color=black, dash pattern=on 1pt off 3pt on 3pt off 3pt]
      table [x=n, y=0.625, col sep=comma] {plot_helstrom_heterodyne.txt};
      \addlegendentry{$\alpha = 5/8$}

      \addplot[color=red, style=solid, forget plot]
      table [x=n, y=hel0.625, col sep=comma] {plot_helstrom_heterodyne.txt};

      \addplot[color=black, dotted]
      table [x=n, y=1.0, col sep=comma] {plot_helstrom_heterodyne.txt};
      \addlegendentry{$\alpha = 1$}

      \addplot[color=red, style=solid, forget plot]
      table [x=n, y=hel1.0, col sep=comma] {plot_helstrom_heterodyne.txt};
    \end{axis}
  \end{tikzpicture}
  \caption{Error probability  $1-P_{c,{\rm E\&D}}^{\rm{(MEC)}}$ of the Estimate\&Discriminate scheme base on a miscalibrated Dolinar receiver where we use the $n$ training copies of $|\alpha\rangle$ of the input~(\ref{INPUT1sym})
  to estimate the value of $\alpha$ via heterodyne mesurements. The red solid line is the Helstrom bound $1-P_{c,\max}^{\rm{(MED)}}$ from~(\ref{def:P_cDiscriminationexact1alpha}).}\label{fig:plot_dolinar_non_calibrato}
\end{figure}
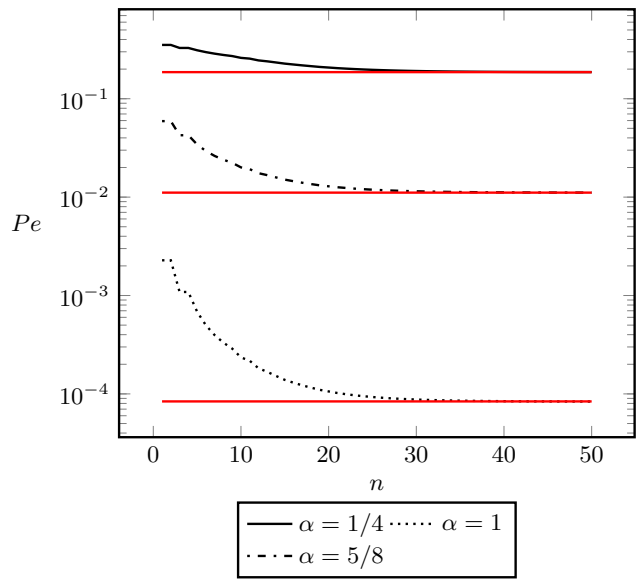

To evaluate the performance  of the scheme  let us first consider what happens when a   Dolinar receiver setup for the discrimination of  $\ket{\beta}, \ket{-\beta},\ \beta \neq \alpha$ is applied to the coherent states $\ket{\alpha}, \ket{-\alpha}$.
For this purpose we can use \cref{eq:differenziale_P_c}, which is still valid, using the optimal displacement evaluated from $\beta$, obtaining a differential equation that can be solved analytically.
As a result we get the following probability of success
\begin{align}
   P_{c,{\rm Dol}}^{(\beta;\alpha)} =\frac{1}{2} + \frac{ \real{\alpha  \beta^*} \left(1-e^{-2 \left(\vert \alpha \vert^2+\vert \beta \vert^2\right)}\right)}{\left(\abs{\alpha}^2+ \abs{\beta}^2\right) \sqrt{1-e^{-4\abs{\alpha}^2 }}},
	\label{eq:P_c_Miscalibrated_Dolinar_receiver}
\end{align}
which, by construction, is upper-bounded by the optimal value $P_{c,\max}^{\rm{(MED)}}$ of Eq.~(\ref{def:P_cDiscriminationexact1alpha}) -- see Fig.~\ref{fig:P_c_Dolinar_Scalibrato}.
The average success probability of the Estimate\&Discriminate approach can now be obtained by
averaging (\ref{eq:P_c_Miscalibrated_Dolinar_receiver}) with respect to probability $P[\beta; \alpha, n]$ of getting a certain
value of $\beta$ from our $n$ copies of $|\alpha\rangle$, i.e.
\begin{align}
   P_{c,{\rm E\&D}}^{\rm{(MEC)}}
   = \int_{\mathbb C^2} \dd \beta \;
     P[\beta; \alpha, n]\;  P_{c,{\rm Dol}}^{(\beta;\alpha)}\;.
 \end{align}
A plot of this quantity as a function of $n$ for few values of $\alpha$ can be found in \cref{fig:plot_dolinar_non_calibrato} under the assumption that $\beta$ is recovered via heterodyne
detection, so that
\begin{align}
  P[\beta; \alpha, n] = \frac{n}{\pi} \exp\left[-n\abs{\alpha - \beta}^2\right].
\end{align}

\subsubsection{Miscalibrated agnostic-Dolinar receiver}\label{sec:dolinar_copie_finite_stima}
\begin{figure}[t]
  \centering
  \includegraphics[width=0.49\textwidth]{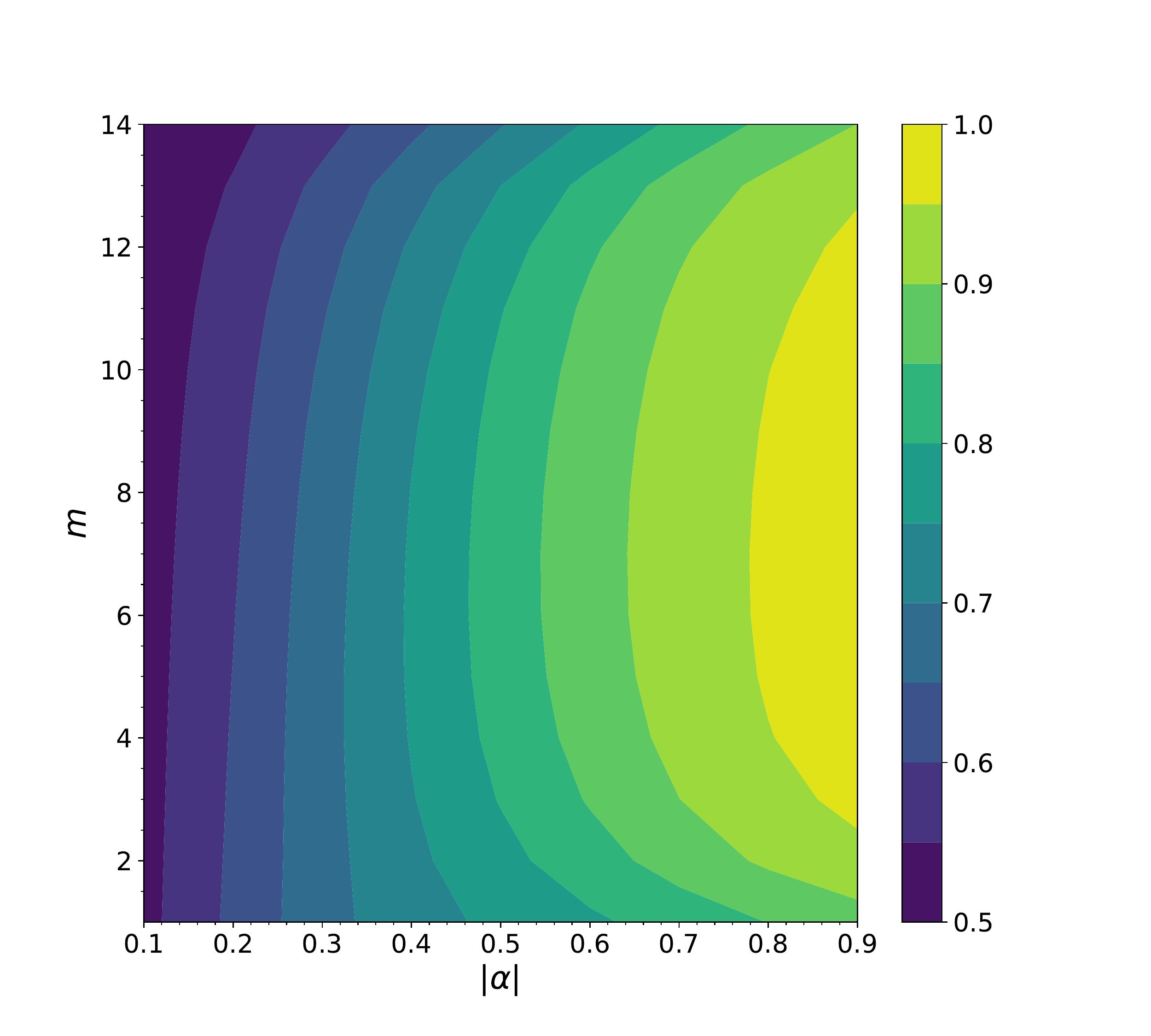}
  \includegraphics[width=0.49\textwidth]{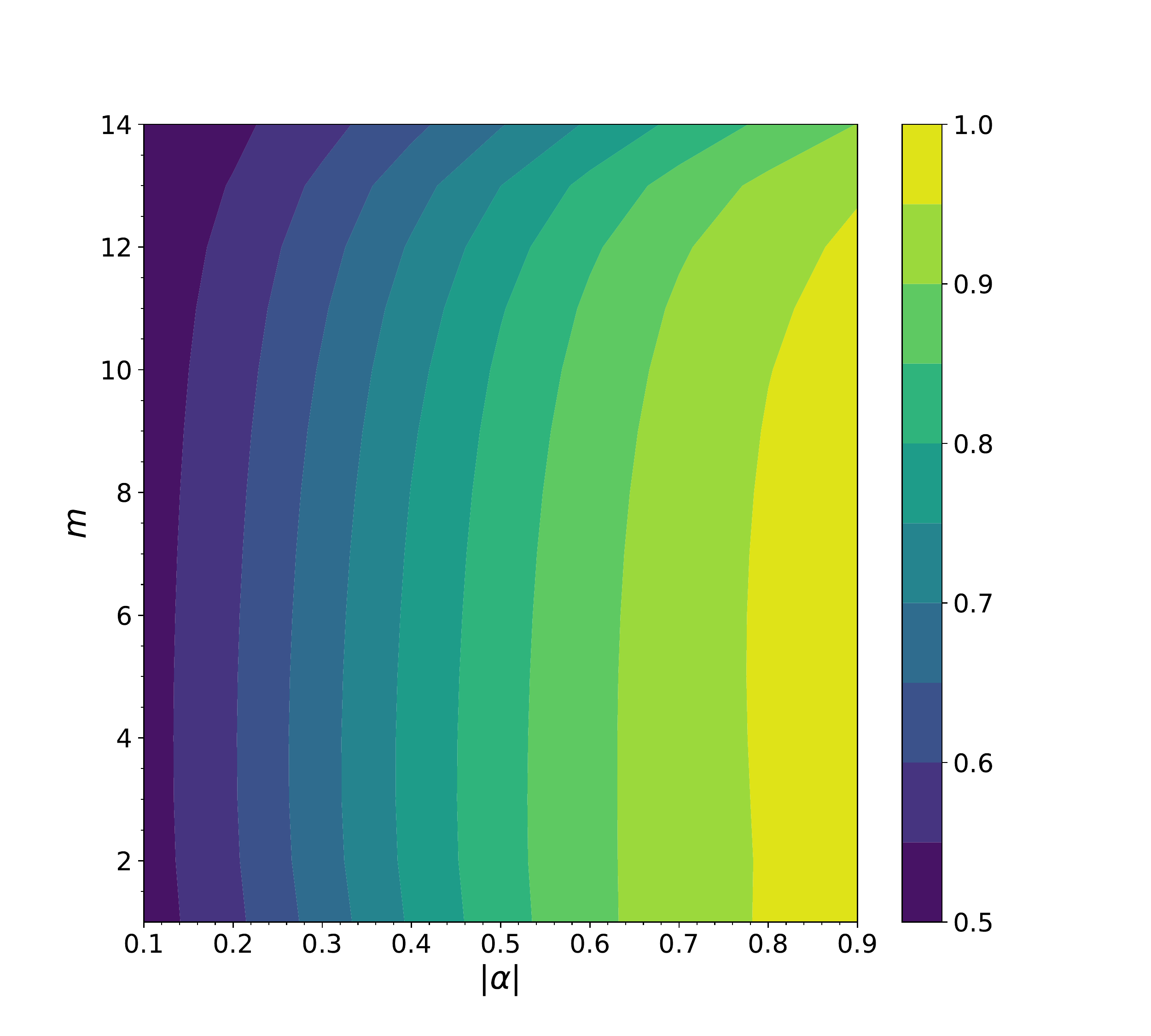}
  \caption{Probability of Correct decision computed as in indicated in Sec.~\ref{sec:dolinar_agnostico_performance} as a function $\alpha$ and the size $m$ of the copies used to estimate it, for a total number of copies $n = 15$. The estimation of $\abs{\alpha}$ is performed via photon--counting in the upper plot and with heterodyne detection in the lower one.}\label{fig:finite_copy_set_division}
\end{figure}

\begin{figure}[!th]
  \centering
	\subfloat[$n=4$\label{fig:confronto_definitivo_n4}]{
		\begin{tikzpicture}
      \begin{axis}[
        xlabel={$\abs{\alpha}$},
        y label style={rotate=-90},
        ylabel={$Pe$},
        ymode=log,
        line width=.9pt,
        transpose legend,
        legend columns=4,
        legend style={at={(0.1,0.1)}, anchor=south west},
        cycle multi list={%
          color list\nextlist
          [2 of]mark list
        }]
        ]
				\addplot[color=blue, dotted]
        table [x=alpha, y=ph_4, col sep=comma] {plot_success_fixed_all__parsed.txt};
        \addlegendentry{Photon-counting}

        \addplot[color=orange, dashed]
        table [x=alpha, y=het_4, col sep=comma] {plot_success_fixed_all__parsed.txt};
        \addlegendentry{Heterodyne}

        \addplot[color=black, dashdotted]
        table [x=alpha, y=mis_4, col sep=comma] {plot_success_fixed_all__parsed.txt};
        \addlegendentry{Miscal.\ E\&D}

				\addplot[color=red, solid]
        table [x=alpha, y=helstrom, col sep=comma] {plot_success_fixed_all__parsed.txt};
        \addlegendentry{Helstrom}
      \end{axis}
    \end{tikzpicture}
	}%
	\\[2mm]
	\subfloat[$n=8$\label{fig:confronto_definitivo_n8}]{
		\begin{tikzpicture}
      \begin{axis}[
        xlabel={$\abs{\alpha}$},
        y label style={rotate=-90},
        ylabel={$Pe$},
        ymode=log,
        line width=.9pt,
        transpose legend,
        legend columns=4,
        legend style={at={(0.1, 0.1)}, anchor=south west},
        cycle multi list={%
          color list\nextlist
          [2 of]mark list
        }]
        ]

				\addplot[color=blue, dotted]
        table [x=alpha, y=ph_8, col sep=comma] {plot_success_fixed_all__parsed.txt};
        \addlegendentry{Photon-counting}

        \addplot[color=orange, dashed]
        table [x=alpha, y=het_8, col sep=comma] {plot_success_fixed_all__parsed.txt};
        \addlegendentry{Heterodyne}

        \addplot[color=black, dashdotted]
        table [x=alpha, y=mis_8, col sep=comma] {plot_success_fixed_all__parsed.txt};
        \addlegendentry{Miscal.\ E\&D}

				\addplot[color=red, solid]
        table [x=alpha, y=helstrom, col sep=comma] {plot_success_fixed_all__parsed.txt};
        \addlegendentry{Helstrom}
      \end{axis}
    \end{tikzpicture}
	}
  \caption{Error probability as a function of $\alpha$ for different classifiers, for size $n=4$ (panel a) and $n=8$ (panel b) of the training set. The red solid line is the Helstrom bound~(\ref{def:P_cDiscriminationexact1alpha}) while the agnostic-Dolinar receiver employing photon counting and heterodyne measurements are depicted with a blue dotted line and orange dashed line respectively. The miscalibrated E\&D line is relative to the Estimate\&Discriminate procedure based on full estimation with heterodyne detection and the use of a conventional (miscalibrated)  Dolinar receiver.}\label{fig:confronto_definitivo}
\end{figure}
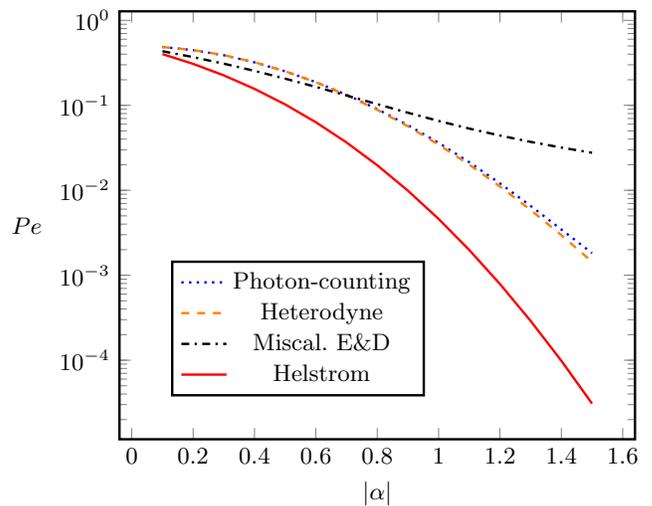
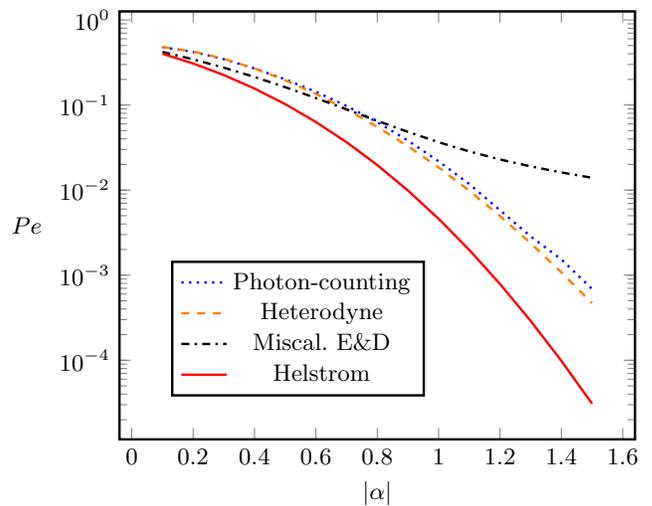

In this section, we study the performance of our classifier in the two-step procedure where we split our set of states into two different sets. The first one, of size $m$, is used to obtain an estimate of the value of $\abs{\alpha}$, while the second one of size $n - m$ is used as input for the agnostic Dolinar receiver described in \cref{sec:dolinar_agnostico}. In the upper plot of \cref{fig:finite_copy_set_division}, the dependence on the size of the estimating set $m$ is studied in case of $n = 15$ with a photon--counting estimator. Other values of $n$ show a similar trend. The same setting, but with heterodyne-detection estimation, is plotted in the same Figure, in the lower panel.
As can be seen from the figure, the optimal choice of $m$ depends on the value of $|\alpha|$, but the optimal value belongs to a big plateau that allows us to ignore this dependence
without losing too much performance. For this reason, we can choose \emph{apriori} the value of $m$ for each $n$, independent from $\alpha$, looking at the plateau in the former figures. With this choice, we can finally compare our results with the Estimate\&Discriminate strategy of Sec.~\ref{sec:dolinar_gaussiano} where all the training copies were used to estimate $\alpha$ with heterodyne detection and then a miscalibrated original Dolinar receiver was employed. These results are summarized in \cref{fig:confronto_definitivo}. The red solid line is the Helstrom bound, while the black lines are the Estimate\&Discriminate performance for $n = 4$ and $n = 8$. The orange and blue lines correspond to photon-counting and heterodyne detection respectively. We can clearly see a divergence in the optimal performance and the Estimate\&Discriminate procedure, due to the difference in the concavity of the two plots. This does not happen with our strategy that remains close to the optimal bound.
For low values of the distance between the states, the performance of the Estimate\&Discriminate is slightly better than our, and this is due to our \emph{apriori} choice of $m$, namely $(n = 4 \to m = 2, n = 8 \to m = 3)$, that is near the plateau for high (greater than $0.3$) values of $\abs{\alpha}$ but is not optimal for low values. If we chose the best $m$ for each value of the distance, our performance would be better than the Estimate\&Discriminate procedure, but this cannot be done for the reasons discussed above.

\subsubsection{Performances in the presence of prior on $|\alpha|$}\label{sec:prior}
\begin{figure}[t]
	\centering
	\subfloat[$n=4$\label{fig:plot_lentissimo_n4}]{
		\begin{tikzpicture}
			\begin{axis}[
					xlabel={$x_c$},
					y label style={rotate=-90},
					ylabel={$Pe$},
					ymode=log,
					line width=.9pt,
					transpose legend,
					legend columns=2,
					legend style={at={(0.1,0.1)},anchor=south west},
					cycle multi list={%
						color list\nextlist
						[2 of]mark list
					}]
					]

					\addplot[color=red, solid, forget plot]
					table [x=alpha, y=opt_4, col sep=comma] {comparison_best_and_rice__parsed_2.txt};

					\addplot[color=orange, dashed]
					table [x=alpha, y=het_4, col sep=comma] {comparison_best_and_rice__parsed_1.txt};
					\addlegendentry{Heterodyne}

					\addplot[color=blue, dotted]
					table [x=alpha, y=ph_4, col sep=comma] {comparison_best_and_rice__parsed_1.txt};
					\addlegendentry{Photon-counting}
			\end{axis}
		\end{tikzpicture}
	}%
	\\[2mm]
	\subfloat[$n=8$\label{fig:plot_lentissimo_n8}]{
		\begin{tikzpicture}
			\begin{axis}[
					xlabel={$x_c$},
					y label style={rotate=-90},
					ylabel={$Pe$},
					ymode=log,
					line width=.9pt,
					transpose legend,
					legend columns=2,
					legend style={at={(0.1,0.1)},anchor=south west},
					cycle multi list={%
						color list\nextlist
						[2 of]mark list
					}]
					]

					\addplot[color=red, solid, forget plot]
					table [x=alpha, y=opt_8, col sep=comma] {comparison_best_and_rice__parsed_2.txt};

					\addplot[color=orange, dashed]
					table [x=alpha, y=het_8, col sep=comma] {comparison_best_and_rice__parsed_1.txt};
					\addlegendentry{Heterodyne}

					\addplot[color=blue, dotted]
					table [x=alpha, y=ph_8, col sep=comma] {comparison_best_and_rice__parsed_1.txt};
					\addlegendentry{Photon-counting}
			\end{axis}
		\end{tikzpicture}
	}
  \caption{Error probability comparison given a prior Rice distribution \cref{eq:rice_distribution} on $\abs{\alpha}$ with $\sigma=0.1$ as a function of $x_c$, for $n=4$ (panel a) and $n=8$ (panel b). The red solid lines represents the associated optimal bound for phase insensitive schemes in \cref{eq:error_phase_invariant}. The performance of the  agnostic Dolinar receiver are plotted for heterodyne (dashed orange) and photon-counting (blue dotted) measurements.}
  \label{fig:plot_lentissimo}
\end{figure}
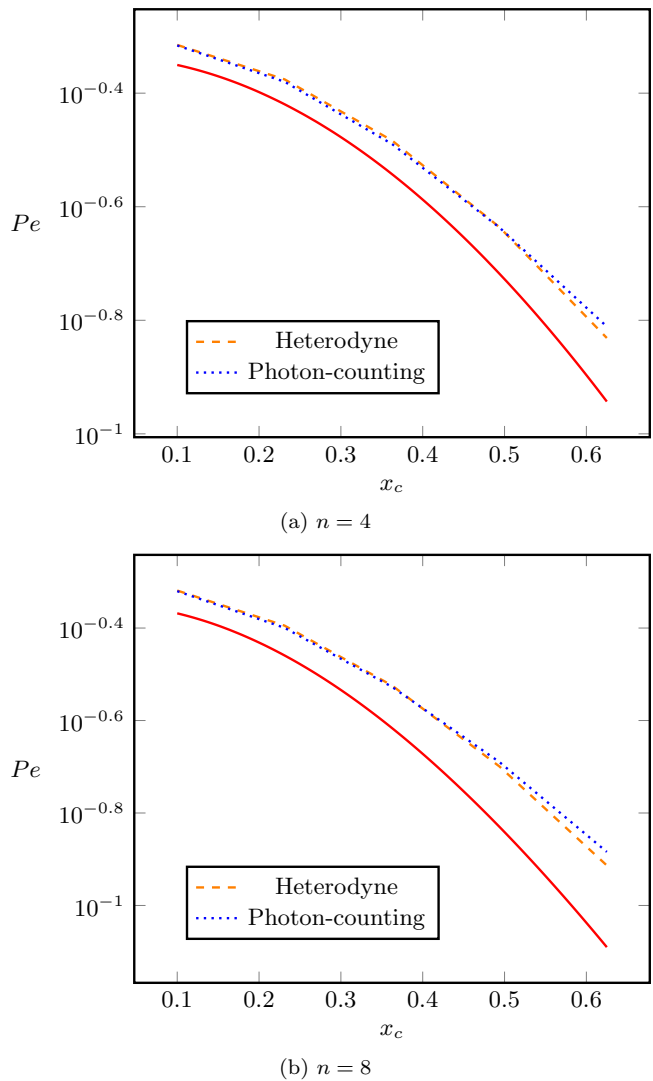

In this section we analyze the agnostic Dolinar scheme when we have a prior on the value of $\abs{\alpha}$ but no information on the value of the phase $\arg{\alpha}$.
In this case  we can average the performance of \cref{fig:finite_copy_set_division} to obtain an expected probability of error for each size of the estimating set $m$, and choose the best $m$ for the given prior.
As an example we consider the case in which the prior distribution for  $|\alpha|$ is given by
the  Rice distribution
\begin{align}
  p(\abs{\alpha}; \sigma, x_c) & = \dint_0^{2\pi} \frac{\exp\left[-\frac{(\abs{\alpha}^2) + x_c^2 - 2x_c \abs{\alpha} \cos\theta}{2\sigma^2}\right] }{2\pi \sigma^2} \abs{\alpha} \dd \theta  \nonumber \\
                               &= \frac{\abs{\alpha}}{\sigma^2} \exp\left[-\frac{\abs{\alpha}^2 + x_c^2}{2\sigma^2}\right] I_0\left(\frac{\abs{\alpha} x_c}{\sigma^2}\right),
															\label{eq:rice_distribution}
\end{align}
where $I_0$ is the modified Bessel function of the first kind.
The obtained results are reported in \cref{fig:plot_lentissimo}, where the optimal error probability bound given in \cref{eq:error_phase_invariant} is compared with the performance of our scheme, employing both heterodyne detection (blue lines) and photon counting (orange lines) to estimate  $\abs{\alpha}$, for $n=4$ and $n=8$. In this plot, which is evaluated with $\sigma=0.1$ as a function of $x_c$, we can observe that with both measurements the performances remain close to the optimal ones as $x_c$ increases, maintaining the ordering with respect to $n$ (greater gives lower error probability).

\section{Conclusions}\label{sec:conc}

To summarize, here we have introduced the Minimum Error Discrimination and Minimum Error Classification problems of quantum states. The former, a central problem in Quantum Information Theory, assumes classical knowledge of the quantum states to discriminate. The latter, risen with the recent studies on machine learning, trades the classical description with the availability of multiple training copies assigned to the classes of quantum states to distinguish.
In Quantum Optics, in the case of the binary discrimination of coherent states, an apparatus realizing the optimal discrimination is known (Dolinar receiver), while the corresponding for the classification problem is still missing. Between these two scenarios, we identify some intermediate setups with increasing level of classical knowledge: for instance, we can assume that the coherent states have opposite phases, or in addition to that, a prior distribution on the amplitude of the quantum states.
We evaluate an optimal bound for this later problem, leveraging on the fact that the POVM associated with the optimal classifier must be phase invariant on the quantum states defined by the training copies and the state to distinguish. This bound asymptotically approaches the Helstrom limit in the limit of infinite training copies, that is the optimal bound for the discrimination problem.

We extend the Dolinar receiver with an agnostic formulation with and without prior information on the input mean photon number. In the case the prior is unknown, a fraction of the training copies is measured to estimate the mean photon number, either with a heterodyne measurement or via photon counting. The remaining training copies are employed in the classification device. We compare the performances of these schemes with the optimal bound previously evaluated and with a Miscalibrated Estimate\&Discriminate apparatus, where all the training copies are employed in the amplitude estimation, used in a later stage by a Dolinar receiver.
The trend of the schemes employing both heterodyne and photon-counting measurements follows the optimal bound with a clear gap, but over--perform the Estimate\&Discriminate strategy. This confirms and extends the results by~\cite{Sentis2015}, where the behaviour of the Estimate\&Discriminate strategy was evaluated asymptotically in the number of training copies.

As future outlooks, one can narrow the gap with the optimal bound with adaptive strategies that estimate the coherent states amplitudes and perform a partial discrimination at the same time. On the experimental side, the proposed classifiers can in principle be already implemented as they require state--of--the--art devices (beam splitters, phase shifters, photon counters and local laser sources) commonly present in current laboratories.

\subsection*{Acknowledgments}
This work has received support from MIUR via PRIN 2017 (Progetto di Ricerca di Interesse Nazionale): project QUSHIP (2017SRNBRK). The authors would like to thank Matteo Rosati for the useful discussions.

\begin{appendices}
  \appendix

\section{An equivalent description of the Dolinar receiver}\label{SEC:DOLINAREQ}

\begin{figure}[t]
  \centering
  \scalebox{0.255}{\input{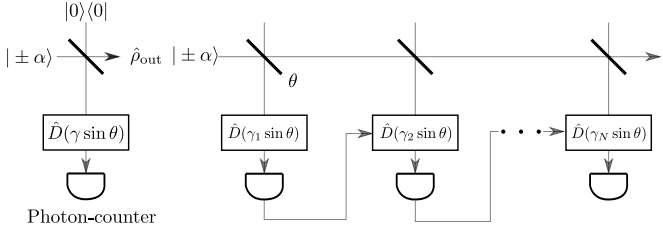}}
  \caption{Discretized description of the Dolinar receiver. The rectangles represents a displacement gate, which is followed by photon-counters, while the diagonal lines represent beam splitters.}\label{fig:dolinar_schema_bs}
\end{figure}

Here we review the alternative formulation of the Dolinar receiver presented in~\cite{dolinar_rivisitato_2}, which is depicted in \cref{fig:dolinar_schema_bs} and comes from the equivalence between a continuous photon--counting process and a sequence of beam splitters and photon--detectors~\cite{quasicontinuous_photocounting}.
The input state comes in the apparatus from the left and goes through a sequence of very similar steps. Each of the diagonal rectangles is a beam splitter of very small reflectivity $\theta \ll 1$. The input state is mixed with the vacuum $\ket{0}$ via this beam splitter, displaced with the displacement gate $\disp{\gamma_k \sin\theta}$, and then undergoes photon counting. The measurement result is used, in addition to the known value of $\alpha$, to decide the next displacement parameter $\gamma_{k+1}$. Then, the discrimination result will simply be the parity of the total number of photons counted. It can be shown that, with the correct choice of $\gamma_k$, and for the number of steps going to infinity, this apparatus tends to the Helstrom bound.

\section{Derivation of~\cref{eq:optimal_error_phase_invariant}}\label{sec:conti_fanizza_analitici_appendice}
Observe that the following identities hold
\begin{eqnarray}
&&\sum_{m=0}^{\infty}\hat{\Pi}_m\ketbra{\sqrt{n}\alpha}{\sqrt{n}\alpha} \otimes \ketbra{\pm\alpha}{\pm\alpha}\hat{\Pi}_m\nonumber\\
&& \qquad =\sum_{m=0}^{\infty} \mathfrak{p}(m;\mu) \ketbra{m,\pm}{\pm, m}, \\
  &&\sum_{m=0}^{\infty} \tracenorm{\Pi_m(\rrho_+-\rrho_-)\Pi_m} = \\
 && \qquad  \sum_{m=0}^{\infty}\mathfrak{p}(m; \mu) \sqrt{1-|\braket{m,+}{}{m,-}|^2}, \nonumber
\end{eqnarray}
which imply \cref{eq:optimal_error_phase_invariant} by noticing that
\begin{align}
  \braket{m,+}{}{m,-}=\left(\frac{n-1}{n+1}\right)^{m}.
\end{align}

\end{appendices}

\bibliography{bibliografia}

\end{document}